\documentclass[sigconf]{acmart}

\AtBeginDocument{%
  }

\setcopyright{acmlicensed}
\copyrightyear{2018}
\acmYear{2018}
\acmDOI{XXXXXXX.XXXXXXX}

\acmConference[ICAIF-24]{Make sure to enter the correct
  conference title from your rights confirmation email}{November 14-16, 2024}{New York City, NY}
\usepackage{algorithm}
\usepackage{algpseudocode}

\begin{document}

\title{Supervised Autoencoders with Fractionally Differentiated Features and Triple Barrier Labelling Enhance Predictions on Noisy Data}
\vspace{1.0cm}
\author{Robert Ślepaczuk}
\affiliation{%
  \institution{University of Warsaw, Faculty of Economic Sciences, Department of Quantitative Finance and Machine Learning, Quantitative Finance Research Group}
  \streetaddress{Ul. Długa 44/50}
  \city{Warsaw}
  \country{Poland}}
\email{rslepaczuk@wne.uw.edu.pl}
\orcid{https://orcid.org/0009-0003-8671-8266}
\author{Bartosz Bieganowski}
\affiliation{%
  \institution{University of Warsaw, Faculty of Economic Sciences, Department of Quantitative Finance and Machine Learning, Quantitative Finance Research Group}
  \streetaddress{Ul. Długa 44/50}
  \city{Warsaw}
  \country{Poland}}
\email{bartosz.bieganowski.office@gmail.com}
\orcid{https://orcid.org/0009-0003-8671-8266}

\vspace{1.0cm}

\renewcommand{\shortauthors}{Ślepaczuk and Bieganowski}

\begin{abstract}
This paper investigates the enhancement of financial time series forecasting with the use of neural networks through supervised autoencoders (SAE), to improve investment strategy performance. Using the Sharpe and Information Ratios, it specifically examines the impact of noise augmentation and triple barrier labeling on risk-adjusted returns. The study focuses on Bitcoin, Litecoin, and Ethereum as the traded assets from January 1, 2016, to April 30, 2022. Findings indicate that supervised autoencoders, with balanced noise augmentation and bottleneck size, significantly boost strategy effectiveness. However, excessive noise and large bottleneck sizes can impair performance.
\end{abstract}




\maketitle
\section*{Introduction}
\addcontentsline{toc}{section}{Introduction}
This research explores an Algorithmic Investment Strategy (AIS) using Supervised Autoencoder - Multi-Layer Perceptron (SAE-MLP) networks, focusing on high-frequency price data. We address two key questions:
\begin{itemize}
\item RQ1. Does data augmentation and denoising via autoencoders combined with triple barrier labeling (TBL) produce better risk-reward metrics than buy-and-hold?
\item RQ2. Does a portfolio of SAE-MLP strategies outperform a similar portfolio of buy-and-hold cryptocurrencies in terms of risk-reward metrics?
\end{itemize}
We study three cryptocurrency pairs from January 1, 2016, to April 31, 2022, using the period from January 1, 2020, to April 31, 2022, for out-of-sample testing. Our approach employs SAE-MLP models for TBL classification using various bar intervals and a walk-forward method.

Previous literature suggests that machine learning models outperform other techniques, particularly statistical approaches (e.g., ARIMA models), especially for non-stationary data (Baranochnikov and Ślepaczuk, 2022). It has been proposed that models perform better when the input value is asset price rather than return (Kijewski and Ślepaczuk, 2020). It has further been suggested that intra-day data-driven strategies should outperform inter-day data-driven strategies. For neural networks, the usage of a supervised autoencoder has been shown to improve generalization performance (Le et al., 2018), including financial time series models (Fons et al., 2020).
In our study, we aim to enhance the research by applying the supervised autoencoder technique to triple barrier labeling to produce a machine-learning algorithmic trading strategy.

The paper is structured as follows. Section 1 details the data and instruments used. Section 2 describes feature engineering methods. Section 3 presents the labeling method and optimization metric. Section 4 discusses data augmentation and model architecture. Section 5 presents the results and summarizes our findings.

\section{Data}

\begin{figure}
    \centering
    \includegraphics[width=1.0\linewidth]{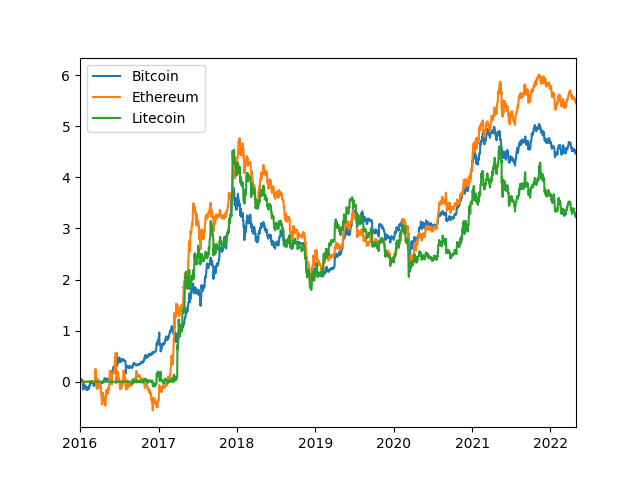}
    \caption{Logarithmic price index for traded currencies.}
    \label{fig:pricepaths}
\end{figure}

This research utilizes three trading time series (Bitcoin, Ethereum, Litecoin) and six feature time series. The data spans from January 2016 to April 2022, with pre-2020 data used exclusively for training. Figure ~\ref{fig:pricepaths} presents the logarithmic growth paths $ln\left(\frac{p_t}{p_0}\right)$ for each of the traded assets.

Each of the time series was obtained from Binance in open high low close (OHLC) format with a 1-minute bar frequency.

Feature time series, sourced from FirstRateData include:
\begin{itemize}
    \item Initial Claims (ICSA): Unemployment data from FRED.
    \item WTI Crude Oil futures: Indicator of economic growth and energy demand.
    \item Henry Hub Natural Gas futures: Another energy demand indicator.
    \item Corn futures: Reflects food-related economic factors.
    \item Gold: Often indicates market uncertainty.
    \item Copper and Aluminum: Reflect industrial activity levels.
\end{itemize}

These features were chosen to represent various economic sectors, mirroring factors traders might consider, spanning from unemployment and general economic activity to market uncertainty and industrial activity levels.

\section{Feature Engineering}

\subsection{Motivation}
Financial time series are notorious for their low signal-to-noise ratios, largely due to arbitrage forces in the market. Standard stationarity transformations, such as integer differentiation, often exacerbate this issue by eliminating valuable historical memory. The use of ARFIMA (Autoregressive Fractionally Integrated Moving Average) models, initially proposed by Granger and Joyeux (1980), offers an insightful perspective. These models allow for fractional differentiation, providing a more nuanced approach to maintaining data stationarity while maximizing memory retention.
\subsection{Fractionally Differentiated Features}
We consider the backshift operator $B$ applied to a time series of a feature ${X_t}$ such that $B^k X_t = X_{t-k}$. The difference between the current and last feature's value can be expressed as $\left(1-B\right)X_t$. For any real number $d$:

\begin{equation}
    \left(1 + x\right)^d = \sum_{k=0}^\infty \binom{d}{k} x^k
\end{equation}

This binomial series can be expanded into a series of weights applied to feature values:

\begin{equation}
    \omega = \left\{ 1, -d, \frac{d(d-1)}{2!}, \frac{d(d-1)(d-1)}{3!}, ..., (-1)^k \prod_{i=0}^{k-1} \frac{d-i}{k!} \right\}
\end{equation}
The weights can be applied to features through multiplication to construct a fractionally differentiated feature. Fractional differencing allows the generalization of differentiation to non-integer orders, capturing long-term memory while ensuring stationarity.
\subsection{Optimal Fractional Differentiation Order}
The fixed-width window fractional differentiation (FFD) method determines the minimum coefficient $d$ that makes the fractionally differentiated series stationary (i.e., passes the Augmented Dickey-Fuller test). If $X_t$ is already stationary, $d = 0$. For a unit root, $d < 1$, while for explosive behaviors, $d > 1$. An especially intriguing scenario is $0 < d < 1$, indicating the series is "slightly non-stationary".

\subsection{Implementation}
Fractional differencing will be applied to features using a walk-forward validation methodology:

Training Segment: Determine optimal $d$ for balance between memory retention and stationarity.
Test Segment: Apply identified $d$ to the succeeding segment for model evaluation.
Rolling Forward: Repeat the process, adapting to evolving series characteristics.

The process can be summarized in the following algorithm:
\begin{algorithm}[H]
\caption{Fractional Feature Differentiation in Walk-Forward Validation}
\small
\begin{algorithmic}[1]
\State Set a range of possible values for $d$ (e.g., from 0 to 1)
\State Set significance level for ADF test (e.g., 1\% )
\State Initiate a dictionary associating each feature with optimal $d$.
\For{each segment pair (train, test)}
\For{each feature}
\State Apply fractional differencing to train segment of the feature at discrete intervals
\State Calculate Augmented Dickey-Fuller (ADF) test statistic and p-value for each d for a feature
\State Choose lowest $d$ such that p-value < significance level.
\State Save feature name and associated optimal $d$ to the dictionary
\State Apply optimal $d$ differencing to both train and test set of the feature
\EndFor
\State Train the model on train segment, evaluate on test set
\EndFor
\end{algorithmic}
\end{algorithm}
This approach ensures features are sufficiently differenced to ensure stationarity, but the memory of the time series is still largely intact, ensuring that our model can benefit from both of the characteristics.

\section{Labelling Methodology}

Most modern approaches to algorithmic trading using machine learning consist of formulating the trading as a classification problem, where the predicted class describes our position (1 - long, -1 - short, 0 - no position) in the market at a given time. In this paper, we are using the triple barrier labeling method. For specified window size $\lambda$, and maximum trade length $n$ minutes, triple barrier labelling for a given time $t$ can be expressed as: 

\begin{equation}
P_t=
    \begin{cases}
      1, & \text{if}\ \max(S_t, ..., S_{t+n}) \geq S_t \cdot (1+\lambda)\\
      -1, & \text{if}\ \min(S_t, ..., S_{t+n}) \leq S_t \cdot (1-\lambda)\\
      0, & \text{otherwise}
    \end{cases}
    \label{eq:tblconditions}
\end{equation}

\begin{figure}
\caption{Triple-barrier-labelling visualization.}
  \centering
  \label{fig:tblpicture}
      \includegraphics[width=1\columnwidth]{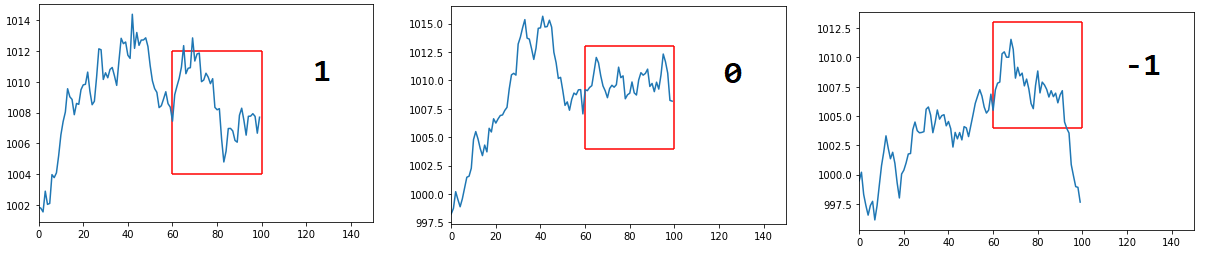}
\end{figure}

Figure ~\ref{fig:tblpicture} represents the three cases visually. In the first case, the upper barrier was exceeded; therefore, we would have preferred to be long at time $t$. In the second case, none of the horizontal barriers was exceeded, so to minimize noise in results, we ideally stay out of the market in this case. In the third case, the lower barrier was exceeded, therefore our preferred position was short. This methodology also assumes in execution, each trade has take-profit and stop-loss set at their respective $S_t \cdot (1+\lambda)$ and $S_t \cdot (1-\lambda)$ levels.\\

\small
\begin{algorithm}[H]
  \caption{Triple-Barrier Labeling - Simple implementation}
  \begin{algorithmic}[1]  
    \State Initialize an empty series \textit{labels} with the same index as \textit{prices}
    \For{each index \textit{idx} in \textit{prices}}
      \State Set \textit{entry\_price} to the price at \textit{idx}
      \State Calculate \textit{profit\_target} as \textit{entry\_price} \( \times (1 + \textit{profit\_taking}) \)
      \State Calculate \textit{stop\_loss\_target} as \textit{entry\_price} \( \times (1 + \textit{stop\_loss}) \)
      \State Set \textit{time\_barrier\_idx} to the minimum of \textit{idx + time\_barrier} and the last index of \textit{prices}
      \For{each price \textit{i} from \textit{idx} to \textit{time\_barrier\_idx}}
        \If{\textit{prices[i]} \(\geq\) \textit{profit\_target}}
          \State Set \textit{labels[idx]} to 1
          \State \textbf{break}
        \ElsIf{\textit{prices[i]} \(\leq\) \textit{stop\_loss\_target}}
          \State Set \textit{labels[idx]} to -1
          \State \textbf{break}
        \ElsIf{\textit{i} is equal to \textit{time\_barrier\_idx}}
          \State Set \textit{labels[idx]} to 0
          \State \textbf{break}
        \EndIf
      \EndFor
    \EndFor
    \State \Return \textit{labels}
  \end{algorithmic}
\end{algorithm}
\begin{center}
\end{center}

It follows from Eq. ~\ref{eq:tblconditions} that on correct (non-zero) prediction, our return on a given trade will always be $\lambda$ (before transactional costs). We define such a case of taking the correct position as a "directly correct" prediction. The "directly incorrect" prediction ($Y_{true} = -Y_{pred}$), on the other hand, will result in a return on the trade of $-\lambda$. On "indirectly incorrect" prediction, for example, $(Y_{true} = 0, Y_{pred}=1)$ the return can be either positive or negative, depending on where the price is at $t+n$, however, will always be within $(-\lambda, \lambda)$. On predicting class 0 we do not open any position therefore our return on the trade will always be zero. Table 2 presents the return distribution for given predicted/true label combinations.

\begin{center}
Table 2. Return on a trade given classification result.\\
\vspace{0.5cm}
\label{tbl:returngivenclassification}
\begin{tabular}{|c | c c c|} 

 \hline
 Pred/True & 1 & 0 & -1 \\ [0.5ex] 
 \hline
 1 & $\lambda$ & $(-\lambda, \lambda)$ &$ -\lambda$ \\ 
 \hline
 0 & 0 & 0 & 0 \\
 \hline
 -1 & $-\lambda$ & $(-\lambda, \lambda)$ & $\lambda$ \\
 \hline
\end{tabular} \\
\vspace{0.5cm}
\end{center} 

\subsection{TBL-Optimized Performance Metric}

To maximize the trading strategy performance, we must introduce an error preference mechanism. After all, missing a profitable trade (type 0 error) will have a lesser effect on our portfolio than entering an incorrect trade (type 1 error). To account for that, we cannot use the accuracy metric in our models. Instead, we have to create a novel, return-maximizing metric. 

We define \textit{directly correct count} as the number of times the model entered the correct position which resulted in the return of $\lambda$. We can similarly define \textit{directly incorrect count} as the number of times the model entered an incorrect position:

\begin{equation}
    DCC = |\{ (Y_{\text{pred}}, Y_{\text{true}}) \in S \mid Y_{\text{pred}} \neq 0 \text{ and } Y_{\text{pred}} = Y_{\text{true}} \}|
\end{equation}
\begin{equation}
    DIC = |\{ (Y_{\text{pred}}, Y_{\text{true}}) \in S \mid Y_{\text{pred}} \neq 0 \text{ and } Y_{\text{pred}} \neq Y_{\text{true}} \}|
\end{equation}

Where $|S|$ is the cardinality of set $S$. It follows from our execution assumptions that cumulative return from trades where $Y_{true} \in (-1, 1)$ can be expressed as:

\begin{equation}
    \Phi = \prod_1^{DCC} (1+\lambda) \cdot \prod_1^{DIC}(1-\lambda) = (1+\lambda)^{DCC} \cdot (1-\lambda)^{DIC} 
\end{equation}

The above equation looks like a good contender for an optimization metric. However, it is important to note that the above equation does not take into account the situation where we enter the trade and the vertical, time-based barrier is reached. We define the number of such trades as \textit{timed exit count} (TEC). We have shown that in these cases the return on the trade will be within $(-\lambda, \lambda)$, however, we cannot assume the average trade return in these situations to be 0. We can therefore introduce a preference mechanism that discourages entering such trades, which also has a much lesser "discouragement magnitude" than for directly incorrect trades. We can do that by constructing the optimization metric as if these trades on average produce a loss, however small it may be:

\begin{equation}
    \Phi = (1+\lambda)^{DCC} \cdot (1-\lambda)^{DIC} \cdot \left(1-\frac{\lambda}{\delta}\right)^{TEC}
\end{equation}

where $\delta > \lambda$. In our study, we set $\delta$ arbitrarily to 20, indicating that twenty timed exits are considered equally undesirable as one direct incorrect classification. Notably, selecting a $\delta$ value lower than $\lambda$ tends to favor trades that have historically led to timed exits. Further research is warranted to explore the potential benefits of this approach for assets characterized by consistent long-term trends, such as the S\&P 500 index.

\section{Model Training Considerations}

\subsection{Data Augmentation}

Data augmentation is a technique that has been pivotal in addressing the issues of overfitting and underrepresentation in machine learning. Originally, its use was most prominent in computer vision problems, where it significantly enhanced the performance of neural networks. In these applications, data augmentation involves making alterations to images in the training dataset to create additional training examples. These alterations can include transformations such as rotating, flipping, scaling, or altering the color balance of images. The augmented dataset thus generated presents a wider variety of scenarios for the model to learn from, which improves its ability to generalize to new, unseen images.

The success of data augmentation in computer vision sparked interest in its potential applicability to other areas of machine learning. In natural language processing (NLP), for example, data augmentation might involve the paraphrasing of sentences or the use of synonyms to expand the dataset. In audio processing, it could involve varying the pitch or adding background noise to sound clips. In tabular data, techniques like feature noise injection or synthetic minority over-sampling are used to enrich the datasets.

Data augmentation has found a valuable place in the domain of time series analysis as well, which is the foundation for many algorithmic trading strategies. Time series data inherently carries the challenge of being sequential, where each point is temporally related to its predecessors and successors. In such a context, traditional data augmentation methods used in computer vision or NLP cannot be directly applied due to the risk of disrupting the time sequence, which is critical to the predictive nature of the data.

In algorithmic trading, the time series data typically consists of historical price movements, volumes, and other financial indicators that are time-dependent. To augment this type of data, the introduction of noise to features based on a fraction of historical feature volatility is an effective technique and is the basis for data augmentation in this paper. This approach preserves the temporal structure of the data while expanding the dataset. By adding noise that is a proportion of the historical volatility, one ensures that the augmented data remains realistic and within the bounds of potential market scenarios.

The noise added is typically Gaussian or drawn from a similar distribution, scaled according to the historical volatility of the feature. For example, if a particular stock has shown volatility of 2\% over a certain period, augmenting the data by adding noise with a standard deviation of 0.2\% (0.1 noise ratio) of the price feature creates new, plausible price paths for the model to learn from. This method of data augmentation helps in creating a more robust algorithmic trading strategy by forcing the model to learn not only from the historical sequence of prices but also from a range of possible price movements that could occur in real market conditions.

The utility of adding noise based on historical feature volatility is twofold. Firstly, it helps in preventing overfitting by ensuring that the model does not learn to anticipate the exact historical sequence of events but rather the underlying patterns that govern market movements. Secondly, it increases the robustness of the algorithmic trading strategy by exposing the model to a wider variety of market conditions during the training phase, enhancing its ability to perform under different market scenarios.

\subsection{Supervised Autoencoder MLP}

\begin{figure}
  \centering
  
  \caption{Supervised autoencoder structure.}
      \includegraphics[width=1.0\columnwidth]{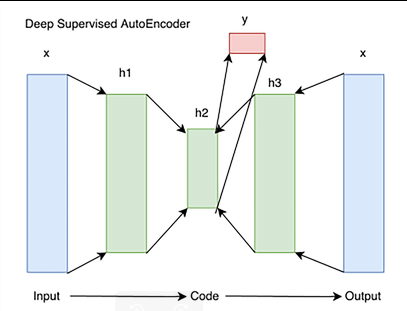}
  \caption*{\footnotesize Source: Esaú Villatoro-Tello, Shantipriya Parida, Sajit 
  Kumar
, Petr Motlicek, Applying Attention-Based Models for Detecting Cognitive Processes and Mental Health Conditions, 2021, \textit{Cognitive Computation}}
\label{fig:sae_structure}
\end{figure}

The neural network type which is examined in detail in this paper is an autoencoder, presented in Figure ~\ref{fig:sae_structure}. An autoencoder is a type of artificial neural network used to learn efficient codings of unlabeled data. The encoding is validated and refined by attempting to regenerate the input from the encoding. The autoencoder learns a representation (encoding) for a set of data, typically for dimensionality reduction, by training the network to ignore insignificant data, leading to finding the most efficient ways to compress passed data. An autoencoder consists of 3 parts: encoder, "code" (also called the bottleneck), and decoder. The encoder compresses the input and produces the code, which is the compressed, denoised data. The decoder then reconstructs the input only using the code. The metric for autoencoder performance is the similarity between reconstructed data and original input.

A supervised autoencoder (SAE) is a variation of an autoencoder that can be used for regression and classification tasks. In SAE, the encoded values are concatenated with the original input and used to train a supervised prediction model on provided labels. The performance metric of SAE is a combination of the accuracy of reconstruction of input data from code (unsupervised part) and the accuracy of predictions using concatenated original data and the code (supervised part). SAE models have shown to have improved generalization performance especially if the data is inherently noisy, which makes it a perfect candidate for a model in algorithmic trading problems.

\subsection{Walk-forward Validation}

In traditional validation approaches, the dataset is split into a training set and a testing set, where the model is trained on the training set and then evaluated on the testing set. However, this approach does not accurately reflect real-world scenarios, where models need to be updated and retrained regularly to adapt to the changing patterns in the data.

\begin{figure}
    \centering
    \caption{Walk-forward validation procedure. The training set, initially expanding, is limited to a 3-period length, therefore shifting instead of expanding since split 4.}
    \label{fig:wfo}
    \includegraphics[width=1.0\columnwidth]{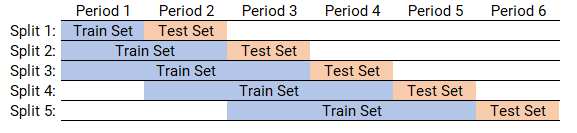}
    \caption*{\footnotesize Note: Final implementation in Python 3.10 using NumPy package}
\end{figure}

Walk-forward validation, presented in Figure ~\ref{fig:wfo}, is a technique that involves dividing the time series dataset into multiple overlapping windows of a constant or expanding size. In each window, a model is trained on the first part of the window (train set) and evaluated on the second one (validation set). This process continues until the entire dataset has been used for training and testing the model. 

Walk-forward validation provides a more accurate estimate of the model's performance in real-world scenarios, where models need to be updated and retrained regularly. It also allows for the detection of changes in the data patterns over time, as the model is evaluated on each overlapping window. Thirdly, it ensures that the model is not overfitted to a specific portion of the dataset, as it is continuously retrained on the latest data. Finally, the constant size window fosters adaptability by requiring the model to perform well across various segments of data that reflect potential shifts in the underlying data-generating process.

The walk-forward method may also be used to tune hyperparameters. A validation period follows the in-sample and is before the out-of-sample in this scenario. The walk-forward model training with the hyperparameter adjustment procedure is analogous to the process described above, with constant window size adjustments to ensure the model remains responsive to the latest data trends.

\subsection{Construction of equity line}

In order to show how SAE-MLPs may be used in algorithmic trading, a straightforward
buy-sell trading strategy is chosen based on whether the instrument price is anticipated to
rise or fall over the next time period. For simplicity, we assume that the orders we place
will not have an effect on the market and that they are executed instantly, at the last
close price. As all three cryptocurrencies are very liquid, this assumption
seems not far from the truth. If our model sends a “buy” signal, the strategy closes out a
short position and takes a long position. If the long position was already taken, it leaves
the position open. If the model sends a “sell” signal the algorithm takes a short position.
To calculate the cumulative unrealized P\&L the following assumptions are used:

\begin{itemize}
    \item The account is opened with \$1.000;
    \item Positions can be opened in any amount, they do not have to be full units;
    \item Transaction costs are calculated for each opening and closing of the position, which
means changing position from short to long will incur double costs. Transaction cost
for all assets amounts to 0.05\%.
\end{itemize}

\subsection{Performance Metrics}

For each strategy and asset, several indicators are computed to evaluate
profitability and performance. When evaluating portfolio performance, it is critical to consider not just the return but also the risk of the strategy. In the study, we utilize performance metrics from Michańków et al. (2022) and Ryś and Ślepaczuk (2018).

\section{Results}

The analysis of the trading strategy utilizing 10-minute, 20-minute, and 30-minute data for Bitcoin reveals intriguing performance characteristics when compared to a buy-and-hold approach. While the buy-and-hold strategy yielded the highest total return (406.02\%) and annualized return (61.63\%), it also exhibited the highest volatility with an annualized standard deviation of 63.31\% and a maximum drawdown of 53.30\%. In contrast, the SAE strategies demonstrated superior risk-adjusted performance, particularly evident in their information ratios (IR) and starred information ratios (IR*). Notably, the 30-minute strategy achieved the highest IR (2.03) and IR* (6.22), coupled with the lowest maximum drawdown (14.02\%) and annualized standard deviation (21.11\%), suggesting enhanced efficiency in capturing market trends while mitigating risk. The 20-minute strategy presented a balanced profile with moderate returns and risk metrics, while the 10-minute strategy, despite higher returns than its longer-interval counterparts, showed increased volatility. These findings indicate that the SAE strategies, especially the 30-minute approach, may offer more favorable risk-adjusted outcomes for Bitcoin trading compared to the buy-and-hold method, despite lower absolute returns.

The analysis of Ethereum trading strategies using 10-minute, 20-minute, and 30-minute data demonstrates remarkably high returns across all approaches, including the buy-and-hold strategy, significantly outperforming the Bitcoin results previously discussed. The buy-and-hold strategy for Ethereum yielded the highest total return (1929.01\%) and annualized return (143.85\%), surpassing all tSAE strategies. However, it also exhibited the highest risk metrics with an annualized standard deviation of 82.85\% and a maximum drawdown of 62.17\%. Among the SAE strategies, the 20-minute approach showed the most favorable balance of risk and return, achieving the highest information ratio (2.64) and starred information ratio (9.51), along with a substantial total return of 1658.93\%. The 30-minute strategy, while showing lower returns, maintained comparable risk-adjusted performance metrics to the 20-minute strategy. The 10-minute strategy, despite producing higher returns than the 30-minute approach, demonstrated increased volatility and drawdown. Notably, all SAE strategies for Ethereum exhibited superior risk-adjusted performance compared to the buy-and-hold method, as evidenced by their higher information ratios and lower maximum drawdowns. This analysis suggests that for Ethereum, the 20-minute strategy may offer the optimal balance between capturing market trends and managing risk, although all strategies, including buy-and-hold, produced exceptional returns in this high-growth cryptocurrency market.

The analysis of Litecoin trading strategies reveals a markedly different performance profile compared to Bitcoin and Ethereum, with SAE strategies significantly outperforming the buy-and-hold approach. Notably, the 30-minute strategy demonstrated the highest total return (382.49\%) and annualized return (59.37\%), surpassing even the 20-minute strategy and substantially outperforming the buy-and-hold method (118.92\% total return). The 30-minute approach also exhibited the most favorable risk-adjusted metrics, with the highest information ratio (2.11) and starred information ratio (4.85), coupled with the lowest annualized standard deviation (28.09\%) and maximum drawdown (25.88\%). The 20-minute strategy showed similar strengths, albeit with slightly higher risk metrics. In contrast, the 10-minute strategy and the buy-and-hold approach both underperformed significantly, with the latter showing particularly high volatility (88.61\% annualized standard deviation) and a severe maximum drawdown (75.21\%). This performance disparity underscores the potential benefits of employing SAE strategies for Litecoin trading, especially those with longer intervals. The superior performance of the 30-minute and 20-minute strategies suggests they may be more effective at capturing Litecoin's price trends while mitigating downside risk, offering a compelling alternative to both shorter-interval trading and passive holding strategies in this market.

The analysis of an equally weighted portfolio comprising Bitcoin, Ethereum, and Litecoin provides valuable insights into the performance of diversified cryptocurrency strategies. Comparing the SAE approaches (10-minute, 20-minute, and 30-minute) to a buy-and-hold strategy reveals interesting dynamics in risk-adjusted returns. The 20-minute strategy achieved the highest total return (619.80\%) and annualized return (79.41\%), slightly outperforming the buy-and-hold approach (595.24\% total return, 77.57\% annualized return). However, the 30-minute strategy demonstrated superior risk-adjusted performance with the highest information ratio (2.56) and starred information ratio (7.63), coupled with the lowest annualized standard deviation (27.37\%) and maximum drawdown (23.55\%). This suggests that the 30-minute strategy most effectively balanced return generation with risk mitigation across the diversified portfolio. The buy-and-hold strategy, while competitive in terms of total return, exhibited significantly higher volatility (72.61\% annualized standard deviation) and maximum drawdown (59.46\%), underscoring the potential benefits of active management through SAE strategies. Notably, all SAE strategies outperformed the buy-and-hold approach in terms of risk-adjusted metrics, with even the 10-minute strategy showing improved risk characteristics despite lower returns. These findings indicate that for a diversified cryptocurrency portfolio, SAE strategies, particularly those with 20-minute and 30-minute intervals, may offer more favorable risk-adjusted outcomes compared to passive holding, effectively capturing market trends while mitigating portfolio-wide risks.

\newpage
\begin{figure}[H]
    \centering
    \includegraphics[width=0.75\linewidth]{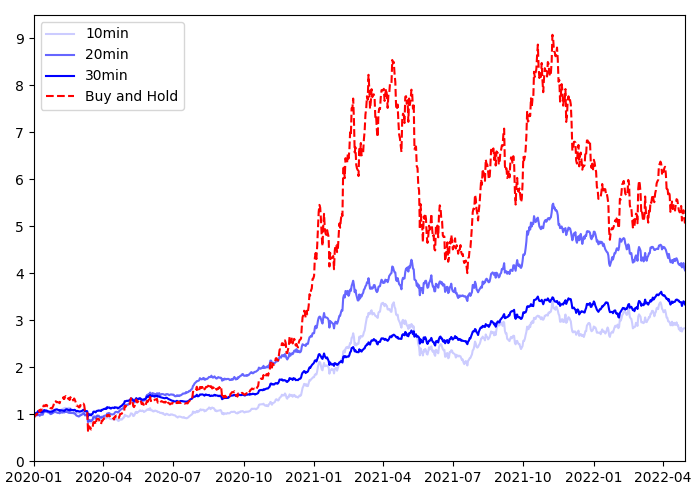}
    \caption{Equity value index for Bitcoin strategies.}
    \label{fig:btc_curves}
\end{figure}

\begin{figure}[H]
    \centering
    \includegraphics[width=0.75\linewidth]{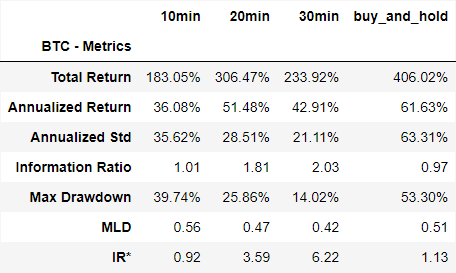}
    \caption{Performance metrics for Bitcoin strategies.}
    \label{fig:btc_metrics}
\end{figure}

\begin{figure}[H]
    \centering
    \includegraphics[width=0.75\linewidth]{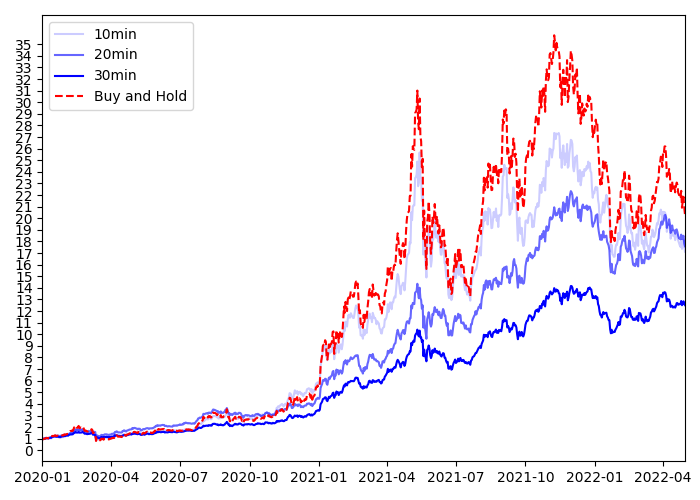}
    \caption{Equity value index for Ethereum strategies.}
    \label{fig:eth_curves}
\end{figure}

\begin{figure}[H]
    \centering
    \includegraphics[width=0.75\linewidth]{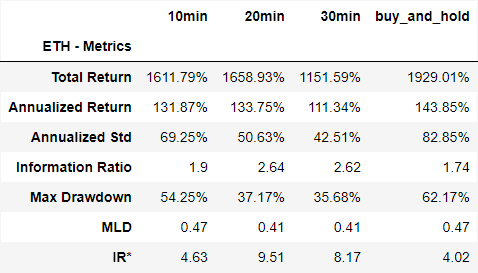}
    \caption{Performance metrics for Ethereum strategies.}
    \label{fig:eth_metrics}
\end{figure}

\begin{figure}[H]
    \centering
    \includegraphics[width=0.75\linewidth]{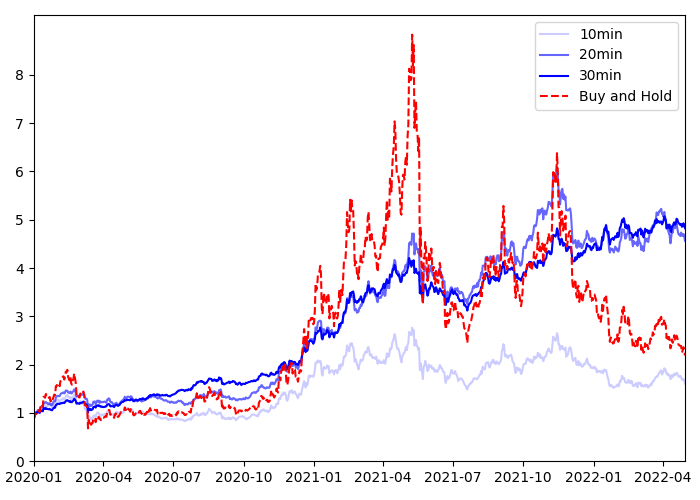}
    \caption{Equity value index for Litecoin strategies.}
    \label{fig:ltc_curves}
\end{figure}

\begin{figure}[H]
    \centering
    \includegraphics[width=0.70\linewidth]{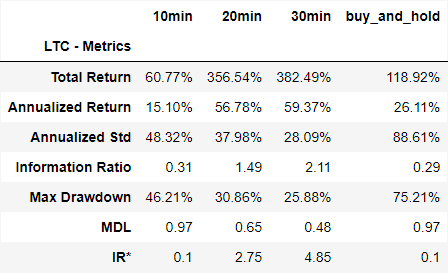}
    \caption{Performance metrics for Litecoin strategies.}
    \label{fig:ltc_metrics}
\end{figure}

\begin{figure}[H]
    \centering
    \includegraphics[width=0.70\linewidth]{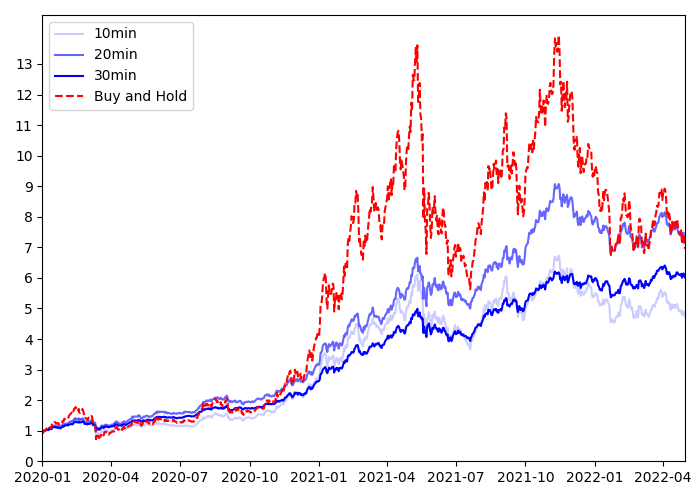}
    \caption{Equity value index for equally weighted portfolio strategies.}
    \label{fig:pfo_curves}
\end{figure}

\begin{figure}[H]
    \centering
    \includegraphics[width=0.70\linewidth]{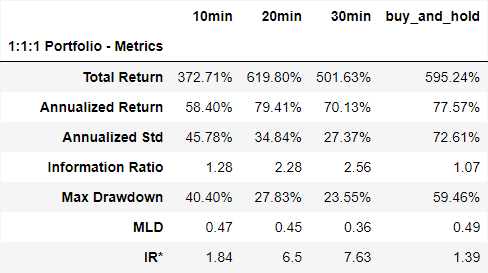}
    \caption{Performance metrics for equally weighted portfolio strategies.}
    \label{fig:pfo_metrics}
\end{figure}

\begin{figure}
    \centering
    \includegraphics[width=1.0\linewidth]{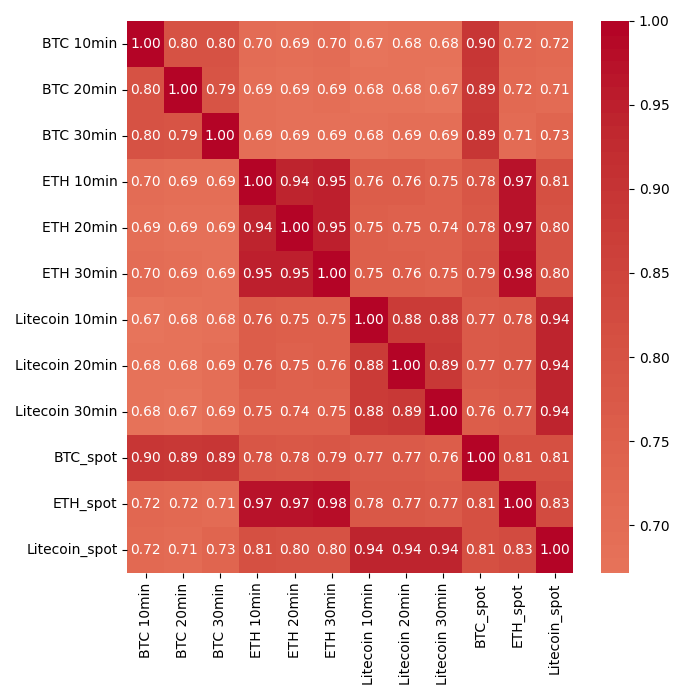}
    \caption{Correlation heatmap for returns of each strategy.}
    \label{fig:corr}
\end{figure}

Although the results look promising we note that Figure ~\ref{fig:corr} demonstrates that the returns of each strategy are highly correlated. We have chosen the currencies that offered the best trade-off between market cap and history length for testing using data from the last day of in-sample data. We note that the test set remained a highly bullish time for these assets, and the results will be affected by that.

\section{Summary}
Our research has demonstrated that the application of Supervised Autoencoder de-noising in combination with Triple Barrier labeling significantly improves algorithmic trading strategies in comparison to buy and hold approaches. Our research has also addressed the following research questions:

\begin{itemize}
\item \textbf{RQ1. Does data augmentation and denoising via autoencoders combined with triple barrier labeling (TBL) produce better risk-reward metrics than buy-and-hold?} - As expressed by information ratio and IR*, the majority of our approaches outperforms the buy-and-hold.
\item \textbf{RQ2. Does a portfolio of SAE-MLP strategies outperform a similar portfolio of buy-and-hold cryptocurrencies in terms of risk-reward metrics?} - Despite the correlation of returns between strategies, combining them into a portfolio outperforms the buy-end-hold equal-weight portfolio.
\end{itemize}

Despite the promising results, the study is subject to certain limitations. Firstly, the findings are based on historical data, and as such, their predictive power in relation to future performance should be considered with caution, given the volatile and evolving nature of financial markets. Secondly, the research does not take into account slippage, which could potentially diminish the net returns of the strategy, especially if dealing in illiquid markets or with large capital. The study also assumes that stop-losses and take profits from triple barrier labeling are executed immediately and perfectly, which is not always the case in the markets.

This paper introduces several new approaches to algorithmic trading. First, it appears to be the first study to apply our specific model architecture in the field of algorithmic trading. This approach differs notably from the traditional models typically seen in this area. Second, while the concept of triple barrier labeling has been previously discussed, our research goes a step further by developing a specialized optimization metric designed explicitly for use with triple barrier labeling. These contributions are significant steps forward in integrating advanced machine-learning techniques into financial trading strategies.

The empirical evidence indicating that algorithmic models can outperform traditional buy-and-hold strategies suggests a necessity for their adoption in asset management to enhance market efficiency and potential investment returns. Consequently, we view it as vital for regulators to craft policies that facilitate the ethical integration of these models, ensuring market fairness and stability while mitigating systemic risks. Institutional investors and fund managers are encouraged to embrace these advanced strategies, necessitating investments in technology and skilled personnel to maintain competitiveness and uphold their fiduciary responsibilities. This shift towards algorithmic trading is not only a reflection of the potential for improved financial performance but also a movement toward the inevitable modernization of financial market practices.

In terms of further research, we recommend investigating other types of noise and their impacts on the strategy's performance. Additionally, the integration of slippage into the model would provide a more realistic picture of the strategy's net returns. Exploring different architectures for the autoencoder or the integration of other deep learning techniques may also yield interesting insights. The impact of these methods on other types of financial time series data, beyond the one used in this study, would also be a fruitful avenue for future research.


\bibliographystyle{ACM-Reference-Format}

\end{document}